\documentclass[a4paper,pre,showpacs,preprint]{revtex4}
\usepackage{graphicx}
\usepackage{amssymb}

\begin{document}

\title{Aggregation of magnetic holes in a rotating magnetic field}

\author{Jozef \v{C}ern\'{a}k, }

\affiliation{P. J. \v{S}af\'{a}rik University in Ko\v{s}ice, Institute of Physics,
Jesenn\'{a} 5, SK-04000 Ko\v{s}ice, Slovak Republic }

\email{jozef.cernak@upjs.sk}

\author{Geir Helgesen,}

\affiliation{Institute for Energy Technology, Physics Department, NO-2027, Kjeller,
Norway}

\begin{abstract}
We have experimentally investigated field induced aggregation of nonmagnetic
particles confined in a magnetic fluid layer when rotating magnetic
fields were applied. After application of a magnetic field rotating
in the plane of the fluid layer, the single particles start to form
two-dimensional (2D) clusters, like doublets, triangels, and more
complex structures. These clusters aggregated again and again to form
bigger clusters. During this nonequilibrium process, a broad range
of cluster sizes was formed, and the scaling exponents, $z$ and $z'$,
of the number of clusters $N(t)\sim t^{z'}$and average cluster size
$S(t)\sim t^{z}$ were calculated. The process could be characterized
as diffusion limited cluster-cluster aggregation. We have found that
all sizes of clusters that occured during an experiment, fall on a
single curve as the dynamic scaling theory predicts. Hovewer, the
characteristic scaling exponents $z',\: z$ and crossover exponents
$\Delta$ were not universal. A particle tracking method was used
to find the dependence of the diffusion coefficients $D_{s}$ on cluster
size $s$. The cluster motions show features of \textit{\emph{Brownian}}
motion. The average diffusion coefficients $<D_{s}>$ depend on the
cluster sizes $s$ as a power law $<D_{s}>\propto s^{\gamma}$ where
values of $\gamma$ as different as $\gamma=-0.62\pm0.19$ and $\gamma=-2.08\pm0.51$
were found in two of the experiments. 
\end{abstract}


\maketitle

\section{Introduction}

Colloidal aggregation phenomena are intersting subjects of study for
both theoretical and technological reasons. In systems with short
range interactions the main aggregation features are well understood
\citet{Vicsek_k}. Diffusion limited cluster-cluster aggregation (CCA)
model \citet{Meakin,kolb} and dynamic scaling theory \citet{Vicsek}
explain well the scaling properties during aggregation. It was found
that these models, initialy developed for systems with short range
interactions, can be used in systems where dipole-dipole interaction
is dominant, for example aggregation of magnetic microspheres \citet{Hel88,Dom},
aggregation of nanoparticles in magnetic fluid \citet{Cer} and aggregation
of magnetic holes \citet{Cer2004}. These experimental results show
scaling of the significant parameters and features typical of CCA.
On the other hand, the corresponding exponents may deviate slighly
from the known models, and the reasons for this is still not understood. 

Our previous results \citet{Cer2004} served as motivation for this
study. Aggregation of magnetic holes in constant magnetic fields was
interpreted in frames of the CCA model and dynamic scaling theory.
The scaling exponent $z\approx0.42$ for the cluster size dependence
$S(t)\sim t^{z}$ was found for particles of diameters $d=1.9$ and
$4\:\mu\mathrm{m}$. This value of the exponent is slighly lower than
exponent values predicted by theory ($z=0.5$) \citet{Meakin} or
exponents found by computer simulations ($z=0.5$ for isotropic and
$z=0.61$ for anisotropic aggregation) \citet{Miguel}. Under certain
experimental conditions (i.e., particles with larger diameter $d=14\:\mu\mathrm{m}$)
the exponent $z$ was close to or lower than the exponent value corresponding
to a transition from 2D aggregation to 1D aggregation ($z=1/3$).
Based on these optical observations we know that for a constant magnetic
field clusters move in 2D but they grow only in one dimension. 

The constant magnetic fields induced formation of long chains of particles
\citet{Hel88,Cer2004}. To determine correct scaling exponents we
may take into account hydrodynamic effects \citet{Miguel}. We proposed
to apply rotating magnetic fields to ensure quasi isotropic properties
inside the magnetic fluid (MF) sample. In this case clusters can move
in 2D space and can grow as 2D compact objects and thus the hydrodynamic
correction is less important. 

The dynamic properties of a few magnetic holes \citet{Skj} in rotating
magnetic fields show interesting phenomena, for example nonlinear
response of bound pairs of magnetic holes \citet{Hel90}, complex
braid dynamics \citet{Pier}, and equilibrium configurations of rotating
particles without contact between particles \citet{Tous}. In a precessing
magnetic field paramagnetic particles dispersed in a drop of water
self-assemble into two-dimensional viscoelastic small clusters \citet{Tierno}. 

In the present study field induced aggregation of many magnetic holes
has been observed. In Sec. \ref{sec:Microscopic} the experimental
equipment and the methods used are described. The Sec. \ref{sec:Results}
deals with the results concerning the determination of the scaling
exponents and characterization of the diffusion behaviour of individual
clusters by tracking of their motions. In Sec. \ref{sec:Discussion}
we summarize the general features and try to explain the non-universal
scaling exponents. Our conclusions follow in Sec. \ref{sec:Conclusions}.

\section{\label{sec:Microscopic}Microscopic observations}

\begin{figure}
\includegraphics[width=8cm]{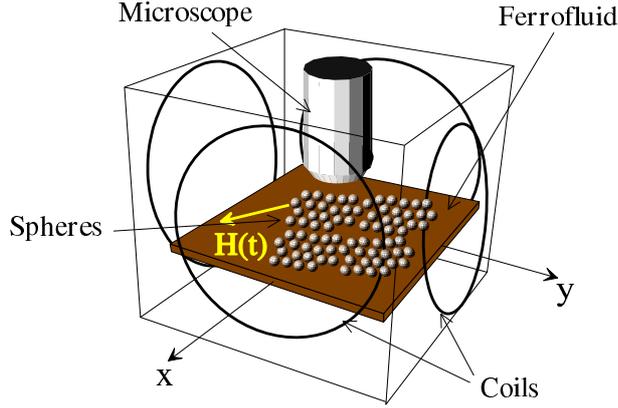}

\caption{\label{fig:1}(colour online) Experimental setup used to study aggregation
of magnetic holes. A rotating magnetic field $\mathbf{H}(t)$ is applied
in the plane of the magnetic fluid. }

\end{figure}

The experimental setup shown in Fig. \ref{fig:1} consists of an optical
microscope (Nikon Optiphot), two pairs of coils, and a carefully prepared
thin layer sample. Alternating curents were supplied to the coils
in order to produce a magnetic field rotating in the horizontal plane
of the sample. Microscopic observations were captured by a CCD camera
(Q-Imaging Micropublisher 5) with resulution $2560\times1920$ pixels. 

The sample size was about $20\times20\:\mathrm{mm^{2}}$. A layer
of magnetic fluid of thickness approximately $50\,\mu$m was confined
between two glass plates and sealed. The kerosene based magnetic fluid
\citet{MF} had the following physical properies: density $\rho=1020\:\mathrm{kgm^{3}}$,
susceptibility $\chi=0.8$, saturation magnetization $M_{s}=20\:\mathrm{mT}$,
and viscosity $\eta=6\times10^{3}\:\mathrm{Nsm^{2}}$. Monodisperse
polystyrene microspheres of diameter $d=3\:\mu\mathrm{m}$ were dispersed
in the MF layer in order to create magnetic holes in presence of magnetic
fields.

\begin{figure}
\includegraphics[width=8cm]{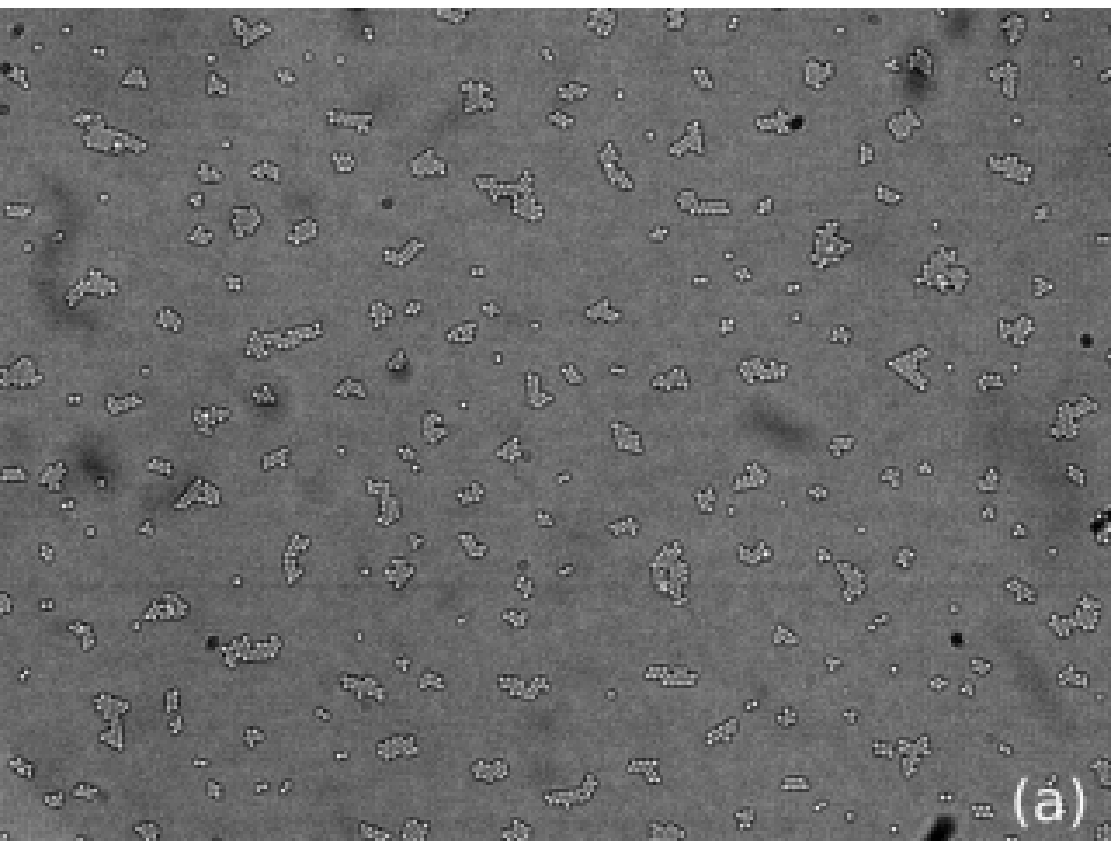}

\includegraphics[width=8cm]{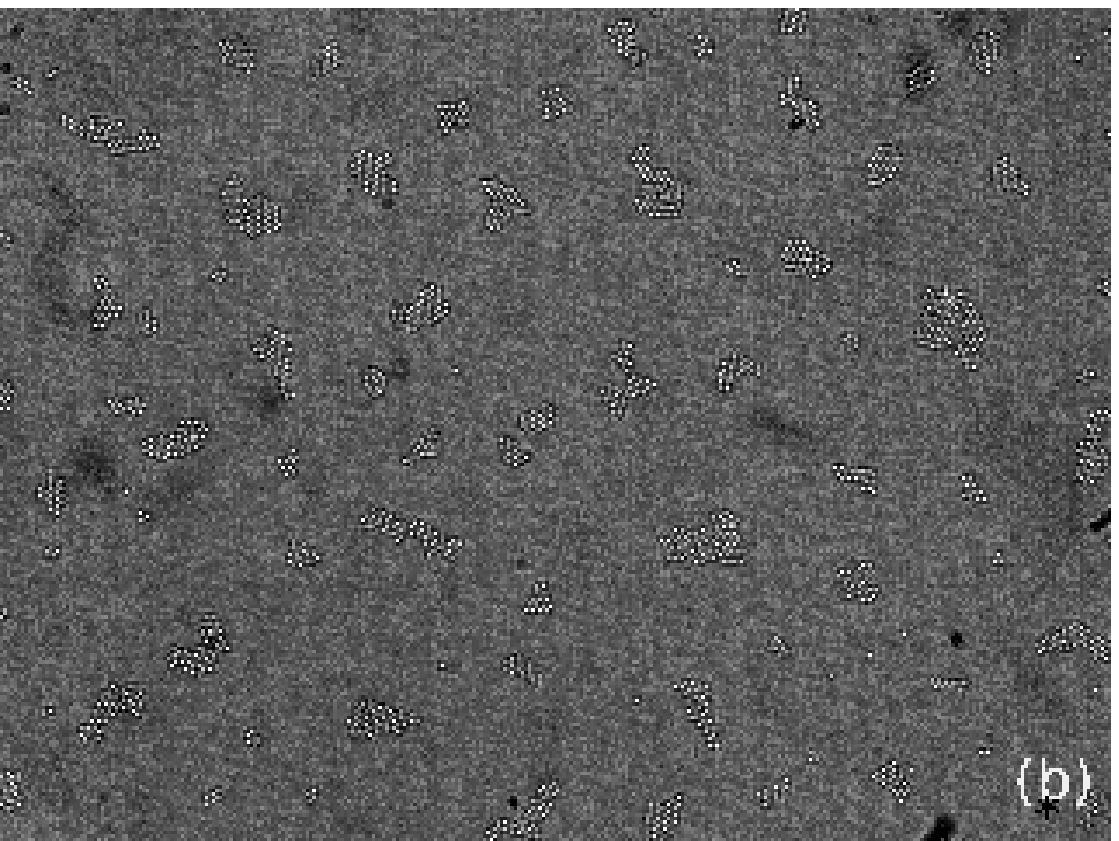}

\includegraphics[width=8cm]{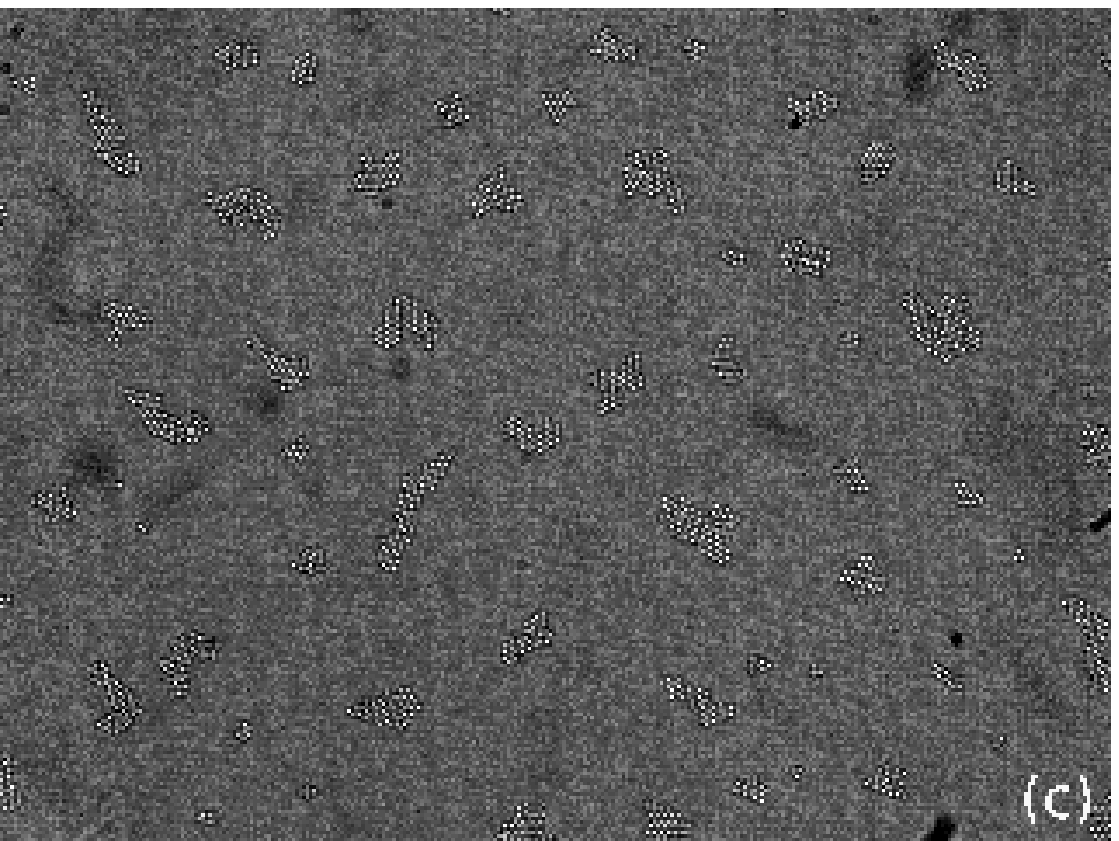}

\caption{\label{fig:2}Optical micrographs of aggregation of nonmagnetic microspheres
with diameter $d=3\mu$m in magnetic fluid at different times: (a)
$t=600$ s, (b) $t=3000$ s, and (c) $t=3400$ s after the magnetic
field was switched on. The applied rotating magnetic field had an
amplitude $|\mathbf{H}|=793$ A$\mathrm{m^{-1}}$ and frequency $f=40$
Hz. The optical view covers a sample area of about $368\:\mu$m$\times$$274\:\mu$m.}

\end{figure}

Without a magnetic field the particles are homogeneously dispersed
in the layer and they can move freely. After some time a very low
fraction of particles may randomly join to other particles and a few
dimers were observed \citet{Skj}. Their volume fraction is very low
in comparison with the volume fraction of single particles. However,
before the application of the rotating magnetic field, a short magnetic
field pulse perpendicular to the sample (coils are not shown in Fig.
\ref{fig:1}) was applied in order to destroy these dimers and to
create a monodisperse initial size distribution of particles. This
initial stage of the experiment is not shown in Fig. \ref{fig:2}. 

The rotating magnetic field $\mathbf{H}(t)=(H_{x},\: H_{y})$ within
the $x-y$ plane had the components: $H_{x}=H\sin(\omega t)$ and
$H_{y}=H\sin(\omega t+\pi/2)$. The amplitude of the magnetic field
was constant $H=793$ A$\mathrm{m^{-1}}$, and angular velocity $\omega=251$
$\mathrm{s^{-1}}$. The temperature during the experiments was $T\doteq293$
K. The effective volume susceptibility including the demagnetization
correction for spherical magnetic holes was $\chi_{eff}=\chi/(1+2\chi/3)=0.63$.
The dimensionless interaction strength parameter \citet{Cer2004}
was $\lambda\doteq90$. Here, $\lambda=U_{max}^{dip}/kT$, where $U_{max}^{dip}$\emph{
}is maximal dipolar energy of two joined dipolar particles, $k$ is
Boltzmann's constant and $T$ is the temperature. Thus, dipole-dipole
interaction among magnetic holes was dominant over the thermal fluctuations. 

The rotating magnetic field induced an aggregation of the microspheres.
The process took place via the joining of single particles into dimers,
trimers and formation of 2D clusters consisting of many particles.
These new clusters aggregate again and formed bigger clusters. A typical
aggregation dynamics is shown in Fig. \ref{fig:2}. A few samples
with approximately the same layer thickness were investigated. The
volume fractions of particles were low, in the range $\phi=0.0014-0.0064$.

In order to analyze the digital images, a C programing language code
and open graphical libraries were used. Several thousands of digital
pictures have been analyzed in a distributed manner in a computational
grid. We have analyzed the motion of individual particles during aggregation
using our own tracking algorithms written in the Python programing
language. The main advantage of the algorithm that was used is that
it can track positions of new clusters which are results of the aggregation.

\section{\label{sec:Results}Results}

Microspheres inside a magnetic fluid layer witout magnetic field behave
as nonmagnetic particles dispersed in fluid. They perform random \textit{\emph{Brownian}}
motion. In this case, aggregation events are rare due to the low particle
concentration. Thus, in the initial stage of the experiments the microsperes
are homogeneously dispersed in the layer of MF and the cluster size
distribution is unimodal. 

After application of the external magnetic field the microspheres
begin to behave as interacting magnetic holes. They have induced magnetic
moments which are oppositely oriented to the external magnetic field.
When the energy of dipole-dipole interaction among two arbitrary spheres
is larger than the thermal energy of the spheres, as quantified by
the dimensionless interaction strength $\lambda\doteq90$ in the present
case, field induced aggregation starts. 

During the aggregation complex motions of microsperes and clusters
consisting of many microspheres were observed. Clusters containing
regularly ordered particles were formed and small irregular clusters
relatively quickly relax to highly ordered structures. Based on the
optical observation the complex modes of motion of microsperes and
clusters may be classified as; i) joining of two clusters together
followed by a very slow relaxation of the microsperes in the new cluster
into a more ordered structure; ii) extremely slowly swivelling of
all clusters in the same direction as the rotating magnetic field,
followed by packing into a compact disk form; and iii) small random
motions of the clusters induced by random forces resulting from interactions
with the local cluster environment.

\begin{figure}
\includegraphics{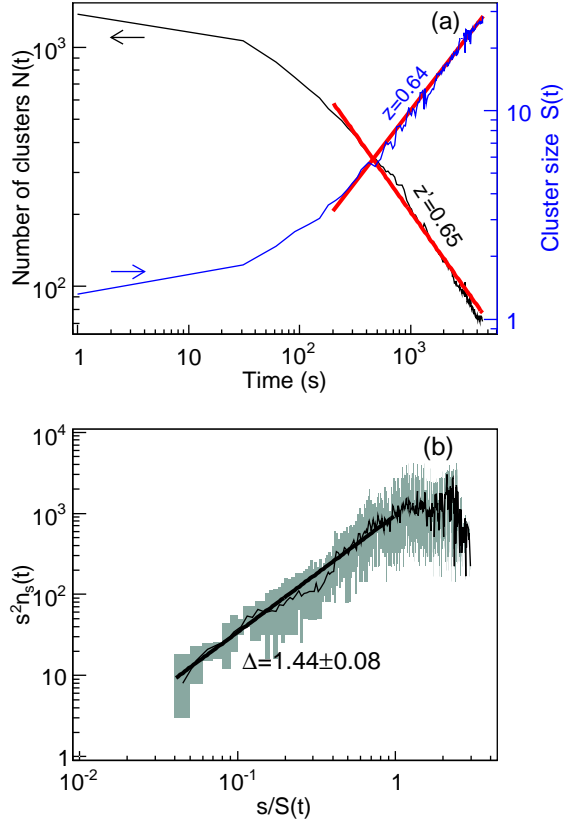}

\caption{\label{fig:3}(color online) (a) The total number of clusters $N(t)$
and the mean (weight average) cluster size $S(t)$ {[}in units of
number of spheres] versus time. (b) The scaling function $g(x)=s^{2}n_{s}(t)$
obtained from the cluster size distributions $n_{s}(t)$ during the
time interval $t=200-3400$ s. }

\end{figure}

We have observed that clusters of all sizes can join together and
form bigger cluster which is the basic feature of cluster-cluster
aggregation. The cluster-cluster aggregation model \citet{Meakin}
predicts the scaling properties of the total number of clusters $N(t)$
and mean cluster size $S(t)$. The total number of clusters is defined
as $N(t)=\sum_{s}n_{s}(t)$ where $n_{s}(t)$ is number of clusters
of size $s$ at time $t$. The mean cluster size $S(t)$ is defined
as:

\begin{equation}
S(t)=\frac{\sum_{s}n_{s}(t)s^{2}}{\sum_{s}n_{s}(t)s}\label{eq:1}\end{equation}
where $s$ is cluster size. In our case the cluster size $s$ is given
by the number of particles which belong to the cluster.

The aggregation process in Fig. \ref{fig:2} was studied in more details.
In Fig. \ref{fig:3} (a) we can see that the total number of clusters
$N(t)$ and mean cluster size $S(t)$ show power law dependencies
$N(t)\thicksim t^{z'}$ and $S(t)\thicksim t^{z}$. The power-law
behaviour was found for the time interval $t=200-3400$ s. The scaling
exponents were determined as $z'=0.65\pm0.01$ and $z=0.64\pm0.01$.

Based on dynamic scaling theory \citet{Vicsek} all number of clusters
$n_{s}(t)$ observed during aggregation can be scaled into a single,
universal curve or scaling function $g(x)$ defined as: 

\begin{equation}
n_{s}(t)\sim s^{-2}g(s/t^{z}).\label{eq:2}\end{equation}
It is expected that $g(x)\sim x^{\Delta}$ for $x\ll1$. 

\begin{figure}
\includegraphics{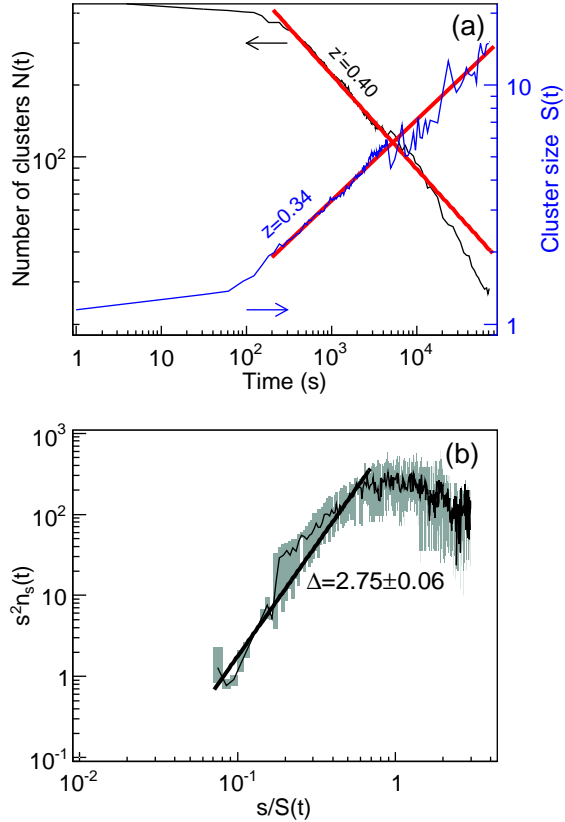}

\caption{\label{fig:4}(color online) (a) The total number of clusters $N(t)$
and the mean (weight average) cluster size $S(t)$ {[}in units of
number of spheres] versus time. (b) The universal scaling function
$g(x)\sim x^{\Delta}$ ($x=s/S(t)$, $x<1$) calculated for the time
intervall $t=200-80000$ s.}

\end{figure}

All the cluster number curves $n_{s}(t)$ during the time interval
$t=200-3400$ s fall onto the single curve shown in Fig. \ref{fig:3}.
From that the characteristic scaling exponent $\Delta=1.44\pm0.08$
was found.

We measured several samples, however, other samples behaved in a different
manner. The scaling exponents $z$, $z'$, and crossover exponent
$\Delta$ were different from the results presented above. Typical
results for a sample that shows a different type of behaviour are
shown in Fig. \ref{fig:4}(a). Here the scaling exponents were found
to be $z'=0.40\pm0.03$ and $z=0.34\pm0.02$. Similarily to the case
discussed above, the cluster numbers $n_{s}(t)$ ($t=200-80000$ s)
that were measured for this sample could be scaled onto a single curve
as shown in Fig. \ref{fig:4} (b), but the scaling exponent $\Delta$
was nearly twice as large is in the former case, $\Delta=2.75$$\pm0.06$.
Also in this case the visible dynamic behaviour was consistent with
diffusion limited cluster-cluster aggregation but with clearly different
scaling exponents from those above.

The results presented in Figs. \ref{fig:3} and \ref{fig:4} show
that the scaling exponents for this system can not be universal. In
order to understand this unexpected result we have investigated the
motions of individual clusters in more detail. 

\begin{figure}
\includegraphics[width=8cm]{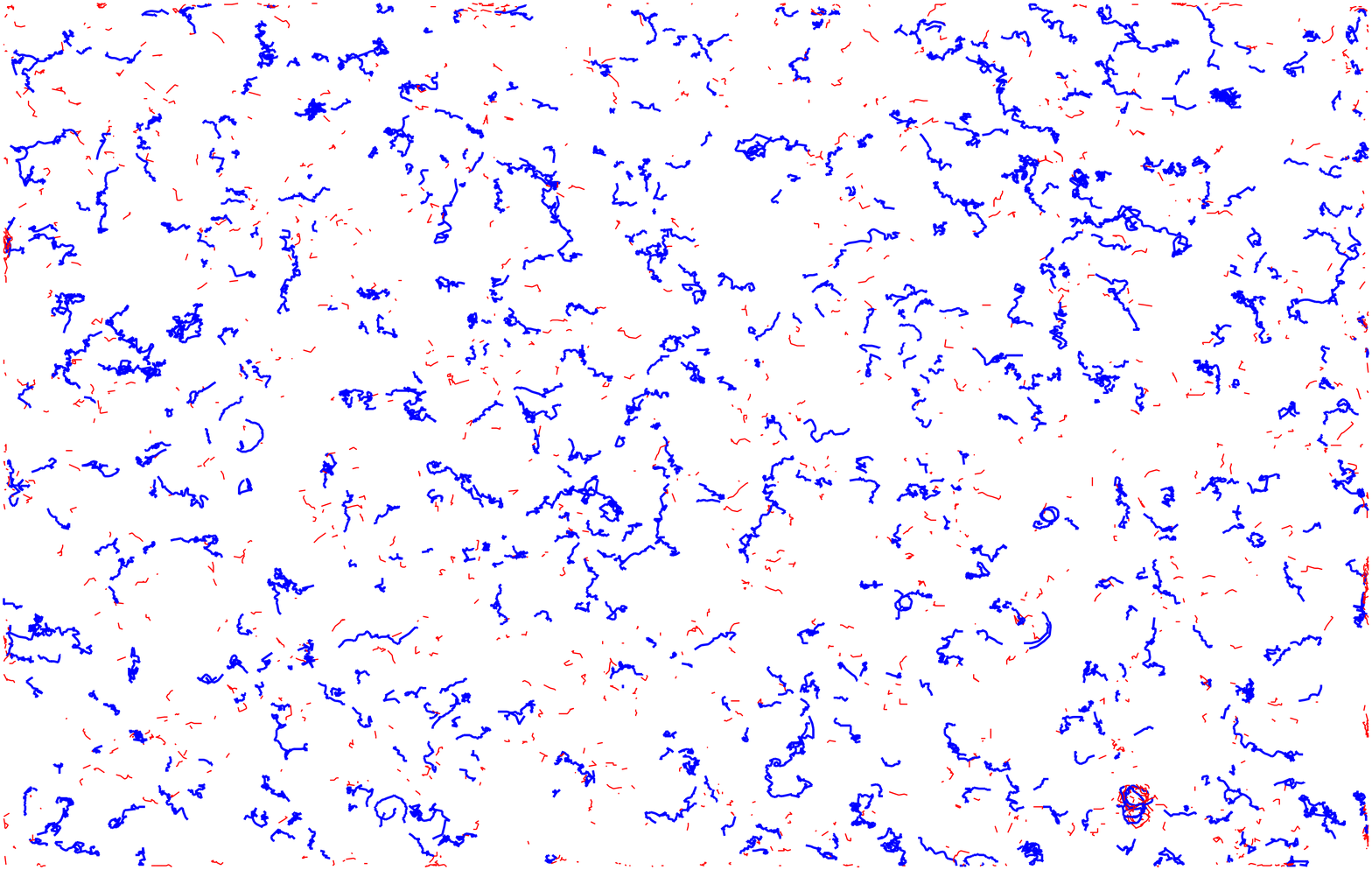}

\caption{\label{fig:5}(color online) Tracks of particles and clusters in the
experiment shown in Figs. \ref{fig:2} and \ref{fig:3} during the
time interval $t=0-3400$ s. Red lines belong to short tracks that
are shorter than 10 time steps (the time step $\Delta t=30\:$s).
Blue tracks are longer than 10 time steps. The sample area is about
$368\:\mu$m$\times$$274\:\mu$m.}

\end{figure}

The complex motion of a cluster was simplified by considering only
the motion of its central mass point. There are effects that can change
the position of the central mass point with nearly no motion of the
cluster as a whole. For example, after joining of two clusters a rearrangement
of particles in the new cluster (see the case i) discussed above)
takes place. We assume that these disturbing changes are smaller than
the influence of random local forces (case iii)) that essentially
contribute to the cluster motions. A very slow rotation of a cluster
(case ii)) does not change the position of the central mass point. 

\begin{figure}
\includegraphics[width=8cm]{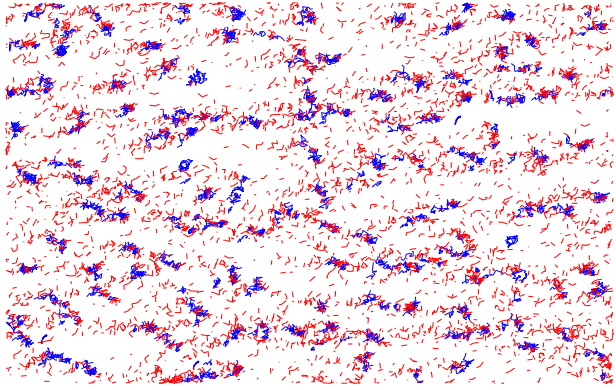}

\caption{\label{fig:6}(color online) Tracks of particles and clusters during
aggregation in the time interval $t=0-3400$ s. Red lines belong to
short tracks that are shorter than 10 time steps (the time step $\Delta t=30\:$s).
Blue tracks are longer than 10 time steps. The sample area is about
$368\:\mu$m$\times$$274\:\mu$m.}

\end{figure}

Cluster tracks shown in Fig. \ref{fig:5} (Fig. \ref{fig:6}) were
determined for experimental data shown in Fig. \ref{fig:3} (Fig.
\ref{fig:4}). We see in Figs. \ref{fig:5} and \ref{fig:6} that
the clusters moved in two directions, the tracks are complex and show
features of \textit{\emph{Brownian}} motion as expected. 

For Brownian particles it is characteristic that their motion is well
described by 

\begin{equation}
<|\mathbf{r}|^{2}>\propto Dt,\label{eq:3}\end{equation}
where $\mathbf{r}$ is the distance vector between an initial position
and the position after time $t$ and $D$ is the diffusion coefficient.
We checked the validity of Eq. \ref{eq:3} for the cluster tracks
and determined the relation $D_{s}\propto<|\mathbf{r}|^{2}>/t$ for
any cluster of the size $s$. At each experiment we analyzed about
$1000$ tracks and found that this equation is valid with a cluster-size
dependent diffusion coefficient $D_{s}$.

\begin{figure}
\includegraphics[width=8cm]{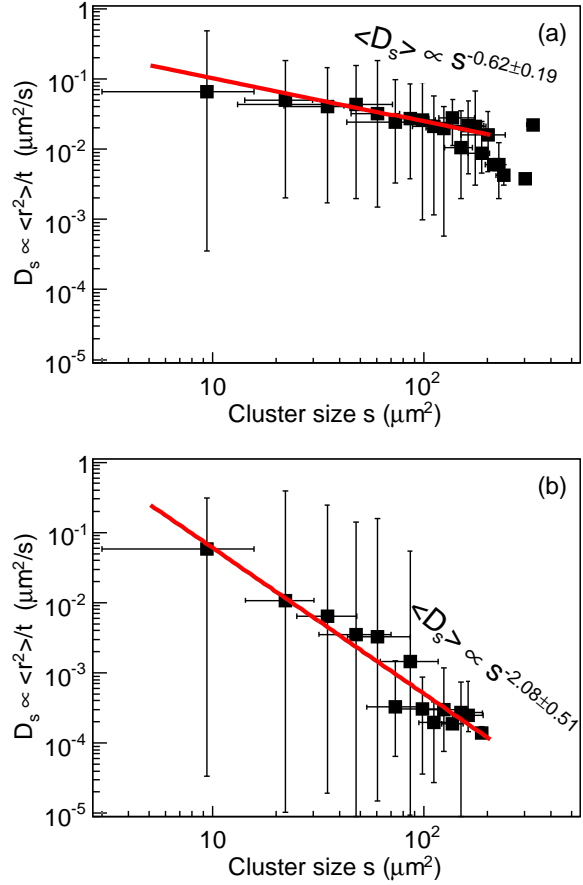}

\caption{\label{fig:7}(color online) The diffusion coefficients vs. cluster
size $s$. (a) For the experimental data shown in Figs. \ref{fig:3}
and \ref{fig:5} the diffusion coefficient follows $<D_{s}>\propto s^{-0.62\pm0.19}$.
(b) For the experimental data shown in Fig. \ref{fig:4} and \ref{fig:6}
the diffusion coefficient scales as $<D_{s}>\propto s^{-2.08\pm0.51}$.}

\end{figure}

We found that $D_{s}$ clearly depended on the cluster size $s$ as
shown in Fig. \ref{fig:7}. For the case in Fig. \ref{fig:7} b) the
values of the diffusion coefficient $D_{s}$ fall in a broad range
covering nearly three decades. These ranges of values of $D_{s}$
are significantly larger than possible errors of measurement. An average
diffusion coefficient $<D_{s}>$ was calculated for any cluster size
$s$ and the results (Fig. \ref{fig:7}) were fitted to a scaling
law $<D_{s}>\propto s^{\gamma}$. For the data presented in Figs.
\ref{fig:3} and \ref{fig:5} the diffusion scaling exponent $\gamma=-0.62\pm0.19$
was found. On the other hand, for the data in Figs. \ref{fig:4} and
\ref{fig:6} the diffusion scaling exponent is clearly higher $\gamma=-2.08\pm0.51$.

\section{\label{sec:Discussion}Discussion}

We have observed that application of a rotating magnetic field on
a 2D magnetic many holes system causes field induced aggregation.
The clusters move and can grow in both dimensions, which is different
from the case of a constant magnetic field where clusters are free
to move in both dimensions but only grow in one dimension as determined
by the external magnetic field. The results show that the system is
in a nonequilibrium state and its characteristic quantities, number
of clusters $N(t)$ and average cluster size $S(t)$, develope and
show scaling according to a cluster-cluster aggregation model \citet{Meakin}
as was shown in Figs. \ref{fig:3}a) and \ref{fig:4}a). Both the
scaling properties and the broad cluster size distributions found,
as well as the existence of a scaling function $g(x)$, are main signatures
of cluster-cluster aggregation and dynamic scaling theory \citet{Vicsek}.

In many cases the cluster-cluster aggregation mechanism leads to formation
of complex, fractal-like objects \citet{Vicsek_k}. However, in the
present case the structure of the aggregates are simpler with a compact
internal organization. In relatively strong magnetic fields, small
clusters are regular 2D objects with well ordered structure of the
particles inside the clusters. Extremely slow cluster rotations and
rearrangement of particles inside the new, bigger clusters have been
observed. As a consequence of these effects the clusters are packed
into regular objects with a nearly close-packed, triangular structure
of the spheres. The cluster diffusion coefficients do not depend on
the direction in which the clusters move, i.e., hydrodynamic corrections
are not important as they are in case of constant magnetic fields
\citet{Miguel}. 

In the basic cluster-cluster aggregation model \citet{Meakin} it
is assumed that the diffusion coefficient is $\gamma=-1$ and the
corresponding scaling exponent is $z=0.5$. We have determined two
distinct values for the diffusion coefficients $\gamma$ and scaling
exponents $z$ as a result of the two clearly different types of behaviour
observed in our experiments. The relationship between $\gamma$ and
$z$ for both values of $\gamma$ follows the equation $z=1/(1-\gamma)$
which has been found in other aggregation models. For $\gamma=-0.62\pm0.19$
($\gamma=-2.08\pm0.51$) we computed $z=0.62$ ($z=0.32$). These
scaling exponents agree well with exponents $z$ determined directly
from the time dependence of $S(t)$, $z=0.64\pm0.01$ and $z=0.34$$\pm0.02,$
respectively. 

Unfortunately, at present we are not able to explain why similar experiments
on approximately the same samples (concentrations, layer thickness
etc.) show scaling exponents with values that come in two clearly
separated ranges and diffusion exponents $\gamma$ which are different
from the expected value $\gamma=-1$. Thus, the scaling exponent $z$
is either clearly lower or higher than the theoretically predicted
value $z=0.5$. 

In an earlier study of a similar system of magnetic holes in a constant
magnetic fields \citet{Cer2004} it was found that for small microspheres
(diameters $d=1.9-4.0\mu$m and interaction strength $\lambda=8-370$)
the scaling exponents $z$ and $z'$ were approximately equal $z\doteq z'$and
typically slightly lower than 0.5: 0.38$\leq z,z'\leq0.54$. However,
for larger particles, $d=14\mu$m ($\lambda=1040-10600$) the values
of $z$ and $z'$ increased with the value of the dimensionless interaction
strength $\lambda$ from about 0.1 to 0.6, and correspondingly the
value of $\Delta$ decreased from above $3.0$ to $\sim1.5$. Thus,
depending on the particle size the scaling exponents changed from
being nearly constant and near the theoretically expected values to
being strongly non-universal. Although the present particles are within
the diameter and $\lambda$ ranges which showed nearly universal behaviour
in Ref. \citet{Cer2004}, the magnetic interactions are very different
(anisotropic in the former and isotropic in the present) and then
the range of $\lambda$ for which the behaviour in non-universal,
seems to be changed. It is unclear why the diffusion conditions, as
quantified by the values of the diffusion coefficients $\gamma$,
were so different in the two typical cases reported here. It may possibly
be related to fine details in the interaction between the microspheres
and the glass plates confining the system. In principle the magnetic
holes should be repelled from the confining walls \citet{Skj} but
if for some unknown reason a small fraction of the particles become
attracted or even loosely attached to the walls, this would slow down
the diffusion as shown by the anomalous value $\gamma\approx-2$ in
one of the analyzed cases. Extremely small values of $<D_{s}>$ could
indicate that some of the particles are trapped in the sample volume
or on the sample glass boundary.

\section{\label{sec:Conclusions}Conclusions}

Diffusion limited cluster-cluster aggregation of magnetic holes has
been induced by a rotating magnetic field. The main features of the
experimental results are well described by a diffusion limited cluster-cluster
aggregation model and dynamic scaling theory. The experimental conditions
were designed in effort to have a well defined model of a low concentrated
many body system where long range interactions are dominant. At present,
the reason why two main aggregation regimes were observed is not clear.
This resulted in scaling exponent values clearly different from those
predicted by theory for systems with short range interactions. This
difference in behaviour was further confirmed by unusual values of
the scaling exponent $\gamma$ of the diffusion coefficient that were
found by cluster tracking. An open question remains: Why do the isotropic,
long-range particle-particle interactions suppress the diffusion regime
where the size-dependence of the diffusion coefficient scales with
$\gamma=-1$? This will hopefully be clarified in future studies. 

\begin{acknowledgments}
The authors thank Arne T. Skjeltorp for many stimulating discussions.
The experimental part of this work has been done at the Institute
for Energy Technology (IFE, Kjeller). J.C. thanks for kind hospitality
at the Physics Department at IFE. Visual data processing was realized
using the results of the projects: Negroid and Know ARC. We acknowledge
financial support from the Slovak Ministry of Education: Grant. No.
6RP/032691/UPJ\v{S}/08. This work was supported by the Slovak Research
and Development Agency under the contract No. RP EU-0006-06.
\end{acknowledgments}


\begin{thebibliography}{10}
\bibitem{Vicsek_k}T. Vicsek, \textit{Fractal Growth Phenomena}, 2nd
ed. (World Scientific Singapore, 1992).

\bibitem{Meakin}P. Meakin, Phys. Rev. Lett. \textbf{51}, 1119 (1983).

\bibitem{kolb}M. Kolb, R. Bottet, and R. Jullien, Phys. Rev. Lett.
\textbf{51}, 1123 (1983).

\bibitem{Vicsek} T. Vicsek and F. Family, Phys. Rev. Lett. \textbf{52},
1669 (1984).

\bibitem{Dom}P. Dom\'{i}nguez-Garcia et al, Phys. Rev. E \textbf{76},
051403 (2007).

\bibitem{Hel88}G. Helgesen \textit{et al}, Phys. Rev. Lett. 61, 1736
(1988).

\bibitem{Cer} J. Cernak, J. Magn. Magn. Mater. \textbf{132}, 258
(1994).

\bibitem{Cer2004} J. \v{C}ern\'{a}k, G. Helgesen, and A. T. Skjeltorp,
Phys. Rev. E \textbf{70},031504 (2004).

\bibitem{Miguel}M. C. Miguel and R. Pastor-Satorras, Phys. Rev. E
\textbf{59}, 826 (1999).

\bibitem{Skj}A. T. Skjeltorp, Phys. Rev. Lett. \textbf{51}, 2306
(1983).

\bibitem{Hel90}G. Helgesen, P. Pieranski, and A. T. Skjeltorp, Phys.
Rev. Lett. \textbf{64}, 1425 (1990). 

\bibitem{Pier} P. Pieranski, S. Clausen, G. Helgesen, and A. T. Skjeltorp,
Phys. Rev. Lett. \textbf{77}, (1996). 

\bibitem{Tous} R. Toussaint \emph{et al}., Phys. Rev. E \textbf{69},
011407 (2004).

\bibitem{Tierno} P. Tierno, R. Muruganathan, and T. M. Fisher, Phys.
Rev. Lett. \textbf{98}, 028301 (2007).

\bibitem{MF} Type EMG 909, produced by Ferrotech, Nashua, New Hampshire.
\end{thebibliography}
\end{document}